# Quantum Phase Transition in the Normal State of High-$T_c$ Cuprates at Optimum Doping

F. F. Balakirev[1], J. B. Betts[1], A. Migliori[1], I. Tsukada[2], Yoichi Ando[3], G. S. Boebinger[4]

[1] *National High Magnetic Field Laboratory, Los Alamos National Laboratory, Los Alamos, NM 87545, USA.*
[2] *Central Research Institute of Electric Power Industry, 2-6-1 Nagasaka, Yokosuka, Kanagawa 240-0196, Japan.*
[3] *Institute of Scientific and Industrial Research, Osaka University, 8-1 Mihogaoka, Ibaraki, Osaka 567-0047, Japan*
[4] *National High Magnetic Field Laboratory, Florida State University, Tallahassee, FL 32310, USA.*

## Abstract

By using a 60 T magnetic field to suppress superconductivity in $La_{2-p}Sr_pCuO_4$, (LSCO) we reveal an anomalous peak in the Hall number, located at optimum doping and developing at temperatures below the zero-field superconducting transition temperature, $T_c$. The anomaly bears a striking resemblance to observations in $Bi_2Sr_{2-x}La_xCuO_{6+\delta}$ (BSLCO) [F. F. Balakirev et al., Nature (London) **424**, 912 (2003)], suggesting a normal state phenomenology common to the cuprates that underlies the high-temperature superconducting phase. The peak is ascribed to the transformation of the "Fermi arcs" into a conventional FS, the signature of a Fermi surface reconstruction associated with a quantum phase transition (QPT) near optimum doping and co-incident with the collapse of the pseudogap state.

The phenomenon of high temperature superconductivity (HTS) occurs in the transition region between an undoped Mott insulator and a Fermi-liquid-like metal with a large Fermi surface (FS) [1]. Many speculate that HTS results from the proximity of a QPT in the underlying normal state both from the theoretical [2-6] and experimental perspective [7-11]. The existence, location, and nature of such a QPT have long been obscured by the superconducting phase; however, normal state behavior at low temperatures emerges once the HTS phase is suppressed with intense magnetic fields [7,8,10].

With initial doping of the parent Mott insulator, even the question of whether the charge carriers form conventional Fermi pockets remains controversial. Angle-resolved photoemission spectroscopy (ARPES) finds well-defined quasiparticles, first near the ($\pi/2$, $\pi/2$) point in the Brillouin zone, then upon further doping, extending along an arc in reciprocal space [12,13]. Debate centers on whether ARPES somehow 'misses' a piece of the Fermi surface, which is proposed to form a 'small pocket' of carriers. The fact that the length of the "Fermi arc" scales with temperature [12], further complicates their interpretation: while some ARPES data suggest that the Fermi arc possibly reduces to a single point upon extrapolation to zero temperature [14], the normal-state electronic specific heat suggests to the contrary [15].

Recent magneto-transport measurements using pulsed magnetic fields as high as 85 tesla have found quantum oscillations in two different underdoped compounds of yttrium-barium-copper-oxide, Ortho-II ordered $YBa_2Cu_3O_{6.5}$ [16] and $YBa_2Cu_4O_8$ [17,18] with effective Cu-O plane doping of 0.10 and 0.12 respectively. These oscillations exhibit the behavior of the well-known Shubnikov-deHaas (SdH) oscillations and thus provide strong evidence of a small and conventional Fermi surface pocket in underdoped cuprates, although its shape and location in the Brillouin zone is still unknown. Subsequently, the observations of quantum oscillations in YBCO have been extended to include magnetization [19].

In the overdoped regime, experiments using a two-axis sample tilt stage in the 45T DC magnet at the National High Magnetic Field Laboratory have measured the angular magneto-resistance oscillations (AMRO) and provided complete three-dimensional mapping of the FS for the single layer thallium cuprate, $Tl_2Ba_2CuO_{6+\delta}$. They reveal a corrugated cylindrical Fermi surface, consistent with the largely two-dimensional nature of the cuprates that describes a large pocket of carriers centered on $(\pi,\pi)$ [1]. The central questions remain: (a) can the evidence of these two distinct states be reconciled, and (b) can the reconciliation provide a framework in which to understand both high-temperature superconductivity and the complex phase diagram of the HTS cuprates?

The most obvious interpretation of the observations to date is that there is a quantum phase transition between the underdoped normal state and the overdoped normal state that is obscured by the intervening superconducting phase. Indeed, a number of HTS models are based on the concept that anomalous electronic properties, including the linear temperature dependence of the resistivity, are governed by the existence of a quantum critical point (QCP). Universal scaling behavior reported in neutron scattering [20], ARPES [21], and infrared spectroscopy [22] is considered as evidence of criticality. Because fluctuations near a critical point can mediate pairing between quasiparticles, quantum fluctuations around the hidden QCP are conjectured to account for HTS itself.

The goal of our transport measurements in high magnetic fields is to search for a hidden QCP in LSCO, and to compare its signature in the Hall effect with that of BSLCO [10].

LSCO thin-film samples were prepared by laser ablation using strontium titanate substrates [23] with eleven values of Sr doping $p$ from 0.08 to 0.22 ($p$=0.08, 0.12, 0.14, 0.16, 0.165, 0.17, 0.175, 0.18, 0.19, 0.20, and 0.22 and the superconducting transition temperature, $T_c$, determined at the onset of resistivity, at 19.4K, 31K, 24K, 28.9K, 28.3K, 30.2K, 28.3K, 27.5K, 24.6K, 17.9K, and 13.6K respectively). All films were characterized by X-ray diffraction and uniformity of low-field magnetotransport properties. All samples show metallic behavior for in-plane transport (i. e. $d\rho_{ab}/dT > 0$) at all temperatures above $T_c$. The samples were patterned in a conventional Hall bar geometry for measurement of the longitudinal resistivity ($\rho_{ab}$) and Hall resistivity ($\rho_{Hall}$). The high magnetic field measurements were performed at National High Magnetic Field Laboratory where a 50 T to 65 T magnetic field was applied perpendicular to the films.

Extremely intense magnetic fields are required to destroy the superconducting state in the cuprate superconductors to reveal the resistive normal state well below $T_c$. In this state, longitudinal resistivity, $\rho_{ab}$, exhibits a metal to insulator (M-I) crossover [24]: that is, the overdoped thin film samples with $p> 0.19$ remain metallic ($d\rho_{ab}/dT > 0$) to $T$~1.5K, the lowest temperature measured, while the underdoped and optimally doped samples display a resistivity minimum and a crossover to insulating behavior ($d\rho_{ab}/dT < 0$) in this temperature range. This transition from an insulating to a metallic normal state occurs at $p$= 0.19 with $\rho_{ab}$ ~0.09 mΩ-cm, a resistivity value similar to one reported for single crystals of LSCO and BSLCO [7,8,25].

Hall experiments reveal more dramatic changes near optimum doping. Figure 1 shows the magnetic field dependence of (a) $\rho_{Hall}$ in superconducting LSCO thin films and (b) the Hall coefficient, $R_H = \rho_{Hall}(H)/H$, for the samples at 20K. Above $T_c$, $\rho_{Hall}(H)$ behaves conventionally (i.e. is largely linear in magnetic field for all fields [26]). Below $T_c$, the low noise level for these measurements enables a determination of the recovery of conventional linear-in-field normal state behavior that extrapolates to the origin (dotted lines) once superconductivity is suppressed by the magnetic field.

The magnetic field dependence of $R_H$ (Fig. 1b) has three dominant attributes: (a) there is no evidence of a magnetic field induced phase transition or sharp change in $R_H$ with magnetic field once superconductivity is suppressed. That is, the suppression of superconductivity with magnetic field seems to reveal the normal state transport, where $R_H$ is expected to be nearly independent of magnetic field; (b) the doping dependence of $R_H$ in the normal state dominates the relatively small magnetic field dependence; and (c) $R_H$ in the normal state generally becomes smaller as the Sr doping, $p$, is increased, the expected behavior for a simple single band metal for which the number of charge carriers is inversely proportional to the magnitude of $R_H$. The most striking feature of Fig. 1b, however, is that *$R_H$ is not a strictly monotonic function of doping* near $p\sim0.17$.

Figure 2 displays the temperature dependence of the high-field $R_H$ extracted from the high-field Hall resistivity measurements (dashed lines in Fig. 1a). We note that $R_H$ is monotonic with doping at high temperatures, but below 50K several $R_H(T)$ curves cross. Figure 2b magnifies the complete data set near optimum doping, evidencing a clear local minimum in the doping dependence for $R_H$ at low temperatures at $p=0.175$.

Figure 3a shows the temperature and doping dependence of $1/R_H$ in the normal state of LSCO. Throughout this paper, we plot $1/R_H$ normalized to the number of holes per copper atom and refer to it as the "Hall number". This provides distinct advantages in communicating the magnitude of $1/R_H$ in familiar and material specific units. It is important to note this is a quantitatively precise notation for all dopings.

At elevated temperatures the Hall number displays distinct monotonic dependence on doping. For very small $p$ high temperature Hall number is nearly equivalent to Sr doping [27], which is consistent with our p=0.08 data in Fig. 3. In the overdoped regime, the apparent divergence of the Hall number with increasing $p$ is most consistent with the gradual change the curvature of the underlying FS from hole-like to electron-like, with eventual zero crossing of $R_H$ in the heavily overdoped samples at $p \sim 0.30$ [23,28].

The salient feature in Fig. 3a is the peak that develops in LSCO near $p$=0.175 at temperatures below ~30K. The peak appears on the background of the otherwise monotonic increase of Hall number with increased doping. The peak (a) occurs precisely at optimum doping, which we define as the doping corresponding to the highest value of $T_c$ in this same set of samples; (b) exhibits a narrow width of $\delta p \sim +/- 0.01$; (c) is largely obscured at zero magnetic fields by the presence of the superconducting phase, i.e. it emerges only below a threshold temperature approximately equal to the maximum value of $T_c$; and (d) appears to have a peak value of roughly one carrier per copper atom. This peak is not unique to LSCO – a strikingly similar feature (Fig. 3b) is seen in another superconducting cuprate, single-layer $Bi_2Sr_{2-x}La_xCuO_{6+\delta}$ [10].

It is unlikely that the peak in the Hall number can be quantitatively interpreted as the actual number of carriers, as this would imply a single-band metal with an isotropic quasiparticle scattering rate around the FS. In fact, the temperature-dependent peak argues strongly against this simplest interpretation. As such, in Fig. 4 we subtract the Hall number at 100K from the entire data set, in effect removing the smooth background and its evolution with doping from the plot of the Hall number in order to focus on the peak in isolation.

ARPES experiments on superconducting LSCO samples (i.e. at zero magnetic field) report FS that are hole-like at p=0.15 and *electron-like* at p=0.22 (insets in Fig. 4c) [29]. It is tempting therefore to ascribe the peak in the Hall number to a topological transition of the underlying FS due to a change in sign of the carriers. However, similar ARPES studies performed on BSLCO argue that the underlying FS does *not* display a change in topology anywhere in the 0.10 to 0.18 doping range (insets of Fig 4d) [30]. Because the Hall anomaly we report near optimum doping in both LSCO and BSLCO argues for a common mechanism, we discuss the anomalous peak in the Hall number in connection with the commonly observed transition from a small-carrier-density metal, characterized by the ARPES Fermi arcs, to a large carrier density metal with a large FS pocket.

The observation of ARPES Fermi arcs remains one of the more unusual phenomena in the underdoped cuprates and their interpretation is still debated. Many have discussed the low number of quasiparticles implied by the observation of the Fermi arcs in connection with the "pseudogap" state, the underdoped regime characterized by a large reduction in carrier density below a characteristic temperature $T^*$. The development of the pseudogap has been linked theoretically to the onset of order, with many candidates for the ordered state having been proposed, including antiferromagnetic correlations [31], a *d*- density wave state [6] and a staggered flux phase [32]. The ordered state is thus often discussed in terms of a reconstruction of the FS [33,34]. Rather than enter the fray among theorists, we discuss our results largely in the context of other experimental observations.

Extrapolations of resistivity data above $T_c$, NMR and specific heat data, taken together from many research groups and many values of carrier doping suggest the collapse of the pseudogap phase at $p \sim 0.19$ [9,35]. The experimental ARPES papers also discuss the difference between Fermi arcs and the large FS in terms of the collapse of the pseudogap state because the Fermi arcs end in the same region of reciprocal space in which there are energy gaps in the pseudogap state [12-14].

How might the collapse of the pseudogap state give rise to a peak in the Hall number? For each of the high-temperature superconducting cuprates, an antiferromagnetic state exists in the undoped ($p = 0$, i.e. half-filled) parent compound [36]. Vestiges of this modulation are retained upon hole doping into the pseudogap regime: neutron scattering finds strong spin correlations with an incommensurate modulation vector near $Q \sim (\pi, \pi)$ in the pseudogap regime [37]. Muon spin relaxation and ac susceptibility in LSCO suggest that this magnetism weakens with increased doping, and does not vanish until the vicinity of optimum doping [38].

The loss of an order parameter in the cuprates upon doping, whatever the specific theoretical perspective, would likely be accompanied by critical fluctuations near the quantum phase transition. Critical fluctuations will occur only within a limited doping range *and temperature range* of the quantum critical point. Critical fluctuations would affect quasiparticle interactions and one would expect transport measurements to reflect fluctuations with time-scales and length-scales longer than the quasiparticle life time and mean free path, respectively.

Critical behavior in the vicinity of a quantum phase transition has long been conjectured as a framework in which to understand both high-temperature superconductivity and the complex phase diagram of the HTS cuprates [2-6]. In fact, it has been suggested that the linear-temperature dependence of the normal-state resistivity might be linked with critical behavior at optimum doping. We discuss the peak in the Hall number in this same context. For the doped, two-dimensional, square-lattice antiferromagnet, we note that criticality has been linked to the nucleation of singular – and attractive – quasiparticle interactions [2]. Although it is beyond the present reach of theory, one can conjecture that singularly attractive interactions would eventually overcome the on-site electronic Coulomb repulsion of the Mott insulator, delocalizing electrons in the vicinity of the quantum critical point. In this picture, the peak in the Hall number would result from the delocalized electrons that would otherwise remain localized due to Mott physics. Although speculative to be sure, we note that a link between the optimal doping for superconductivity and enhanced delocalization due to criticality might provide a natural accounting for the observation of a common temperature scale for the Hall peak anomaly in the normal state and the maximum $T_c$ of the superconducting state in zero magnetic field. Regardless of whether the Hall peak is ultimately understood in terms of critical fluctuations, the observation of the same phenomena in the Hall number of two different hole-doped HTS systems suggests a common quantum phase transition underlying the high-temperature superconducting dome. Given experimental evidence, especially from ARPES, the peak in the Hall number at optimum doping is naturally interpreted as the signature of a transition from the Fermi arc state in the underdoped regime to the large pocket Fermi surface in the overdoped regime. The location of the Hall peak at optimum doping is more robust than the location of the M-I crossover, which is linked to impurity concentration and is observed in the underdoped regime in BSLCO [25], at optimum doping in LSCO single crystals [8], and in the overdoped regime in these LSCO thin films [24].

The work at the National High Magnetic Field Laboratory was supported by the National Science Foundation and DOE Office of Science. YA was supported by KAKENHI 20030004 and 19674002. We thank J.C. Davis, N. Harrison, S. Kivelson, P. Lee, P. Littlewood, R. Macdonald, J. Tranquada, C. Varma and S-C. Zhang for discussions. Correspondence should be addressed to F.F.B. at fedor@lanl.gov.

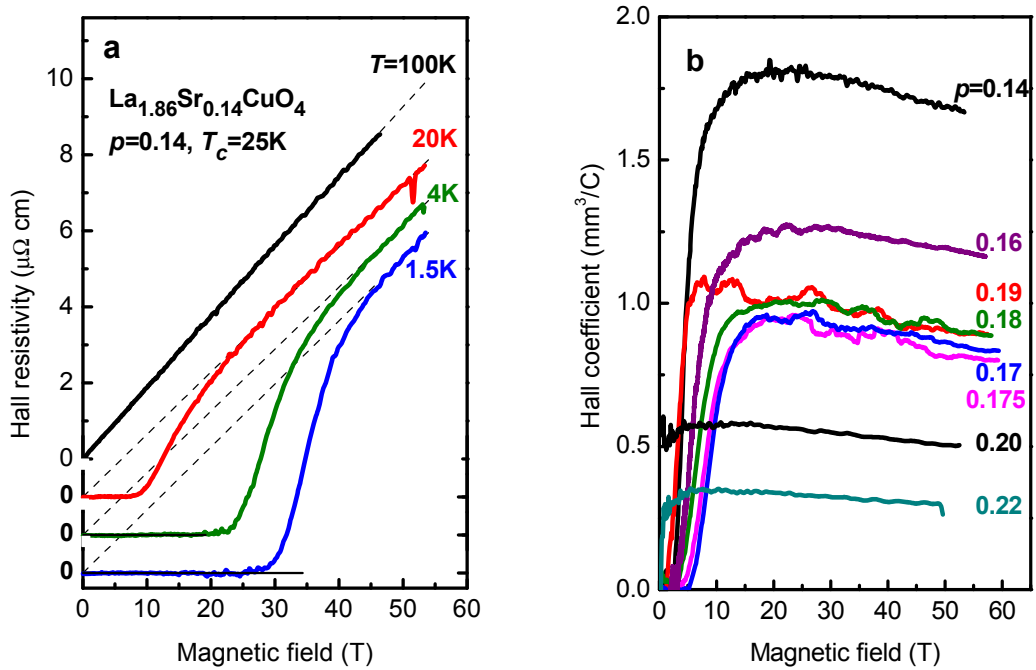

FIG. 1. (a) In-plane Hall resistivity, $\rho_{Hall}$, versus magnetic field for the $La_{1.86}Sr_{0.14}CuO_4$ thin film sample. Dotted lines are an extrapolation of the normal state data, constrained to extrapolate to zero at $H=0$. (b) Hall coefficient, $R_H = \rho_{Hall}(H)/H$, as a function of magnetic field at $T = 20K$ showing the non-monotonic dependence of $R_H$ in the normal state on doping, $p$. Note that the curve for $p=0.175$ lies below the curves for *both* slightly lower doping ($p=0.16$ and $0.17$) and slightly higher doping ($p=0.18$ and $0.19$). For clarity in both (a) and (b), data from many temperatures and doping levels are not plotted.

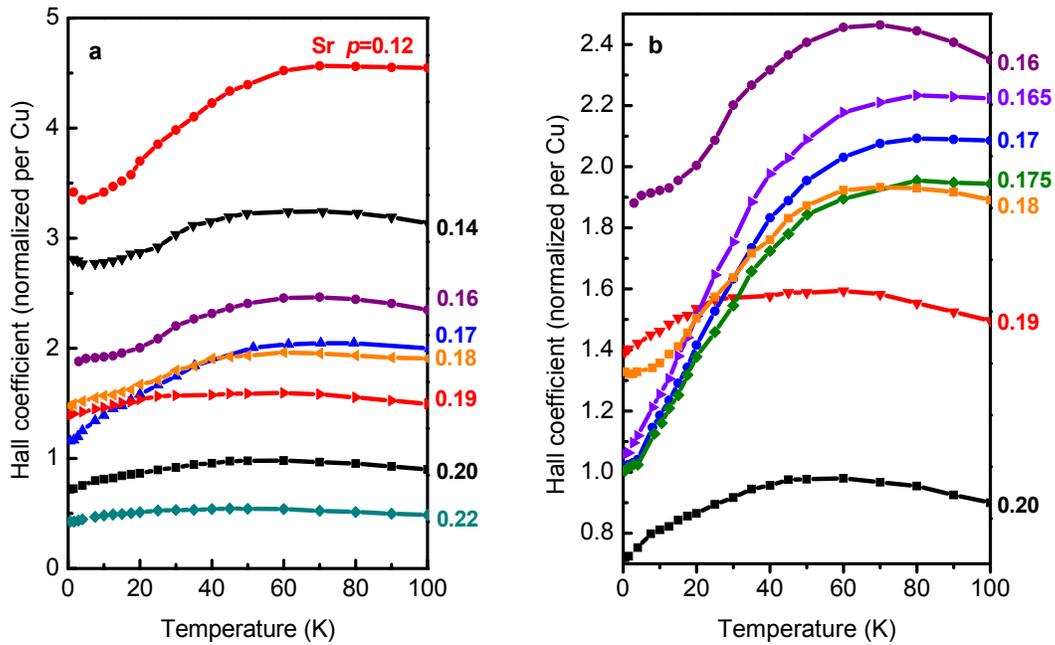

FIG. 2. (a) Temperature dependence of the high-field Hall coefficient, $R_H$, for doping levels $0.12 < p < 0.22$, normalized per Cu atom from the data of Fig 1(b). For clarity, data from doping levels p = 0.08, 0.165 and 0.175 are not plotted. (b) An expanded view of the measurements from all samples with doping levels near optimum doping. While at high temperature $R_H$ doping dependence is monotonic, low temperature data shows a local minimum at $p$=0.175.

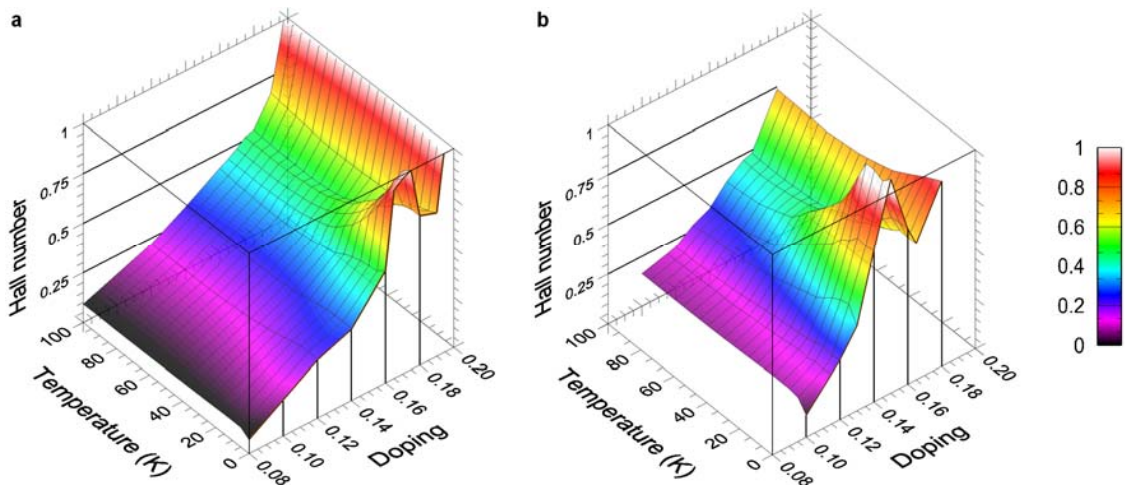

FIG. 3. (a) Doping dependence of the hole-type Hall number (defined as $1/R_H$ and normalized per Cu atom) of $La_{2-p}Sr_pCuO_4$, in which intense magnetic fields have suppressed superconductivity. The grid superimposed on the data indicates the discrete data set from which the surface is deduced. At high temperatures, we find that the Hall number remains relatively low in underdoped samples and that above $p \sim 0.17$ the Hall number increases rapidly with increasing doping. The most striking feature of the Hall number is the cusp at low temperatures that is centered on $p \sim 0.175$, which is the same LSCO sample in which $T_c$ is highest. (b) Hall number in platelets of $Bi_2Sr_{2-x}La_xCuO_{6+\delta}$ single crystals (adapted from Ref. 10), showing a similar low-temperature peak, also occurring near optimum doping and exhibiting a peak amplitude near one carrier per copper atom.

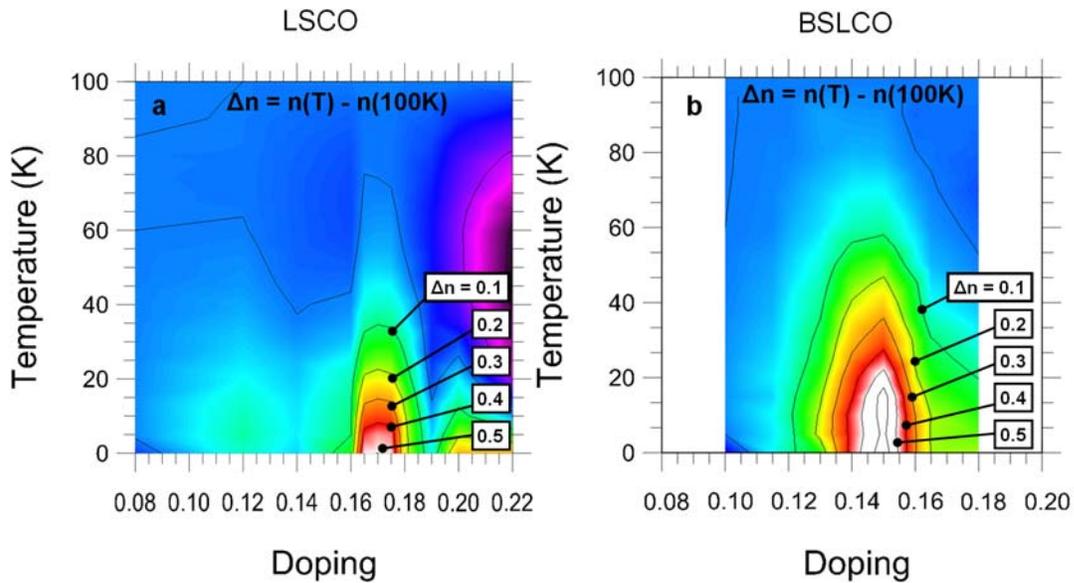

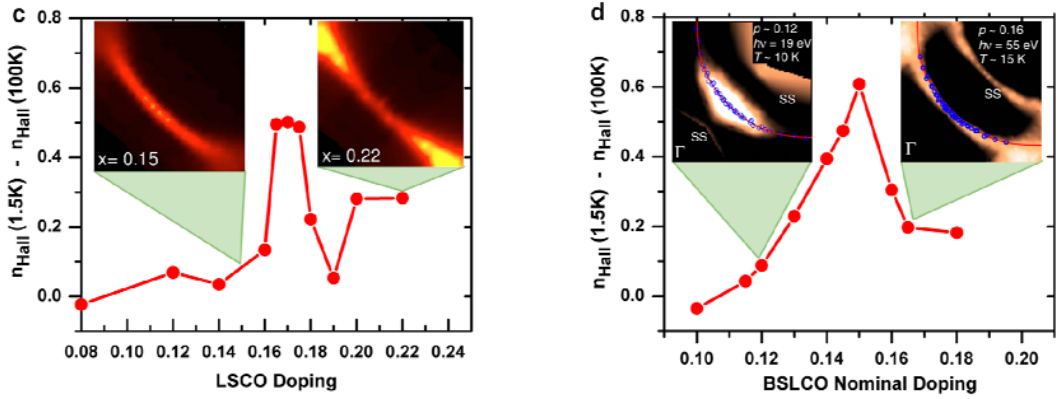

FIG. 4. (a, b) Contour plots of the Hall number variation, $\Delta n = n(T) - n(100K)$, as a function of doping and temperature in (a) LSCO and (b) BSLCO from the data of Fig. 3. (c, d) The low-temperature ($T \sim 1.5K$) value of the Hall number versus doping in (c) LSCO and (d) BSLCO The four insets shows ARPES data for the dopings indicated, reproduced from (c) Ref. 29 and (d) Ref. 30.